\documentstyle[11pt]{article}
\begin{document}
\begin{titlepage}
\vspace{1.5cm}
\title{
{\center \bf Magnetization and overlap distributions
of the ferromagnetic Ising model on the Cayley tree}}
\author{
R. M\'elin $^{(1)}$$^{(2)}$
\thanks{Address from September 1996: International School for Advanced
Studies (SISSA), Via Beirut 2--4, 34014 Trieste, Italy}
\thanks{e--mail: melin@crtbt.polycnrs--gre.fr}
{}\\
{$^{(1)}$ CRTBT--CNRS, 25 Avenue des Martyrs, BP 166X, 38042
Grenoble CEDEX, France}\\
{$^{(2)}$ NEC Research Institute, 4 Independence Way,
Princeton 08540, NJ, USA}\\
{}\\
}
\date{}
\maketitle
\begin{abstract}
\normalsize
We analyze the magnetization and the overlap distributions
on the ferromagnetic Cayley tree.
Two quantities are investigated:
the asymptotic scaling of all
the moments of the magnetization and overlap distributions,
as well as the computation of the fractal dimension of
the magnetization and overlap probability measures.
\end{abstract}
\end{titlepage}

\renewcommand{\thepage}{\arabic{page}}
\setcounter{page}{1}
\baselineskip=17pt plus 0.2pt minus 0.1pt
\section{Introduction}
Spin models on trees were first introduced in the thirties
independently by Bethe and Peierls
as a way to implement mean field theories without long range
order interactions \cite{BP},
even though it was argued recently
\cite{Gujrati}
that in some cases,
Bethe-Peierls calculations are more reliable than
mean field calculations. Even though there is a mean field type
transition for Ising as well as XY spins in the Bethe--Peierls
limit (see section \ref{BPT}) when only the properties
of the central spin are considered and the
boundary is sent to infinity, there is no transition
if one considers the entire Cayley tree: since there are no loops,
the high temperature expansion of the partition function
in a zero magnetic field leads to the same result as
the partition function of the Ising chain,
where it is well-known
that there is only a zero temperature transition.
In the present paper,
we consider the Ising model on the Cayley tree
\begin{equation}
H^{(Ising)} = -J \sum_{\langle i,j \rangle} \sigma_i
\sigma_j
,
\end{equation}
and also the XY model:
\begin{equation}
H^{(XY)} = - J \sum_{\langle i,j \rangle}
{\bf S}_i . {\bf S}_j
.
\end{equation}
The Ising model on the Cayley tree has unusual
properties: it was shown
in \cite{X} that the Cayley tree has a phase transition
of continous order. It was also shown in \cite{Y}
that the susceptibility is infinite below a temperature
$T'$ (that we will recover in the core of the paper).
The aim of the present paper is a detailed investigation
of the magnetization and overlap distributions of the
Ising and XY model on the Cayley tree.

The trees that are considered here are the so--called
{\it half--space--trees},
that is recursive trees such that the ancestor
has two neighbors, the bulk sites have three neighbors and
the leaves have only one neighbor. See figure \ref{Fig1}
for a picture of a $5$--half--space--tree.
We have shown in \cite{Melin} that below a cross--over temperature
scale $J/\ln{n}$, with $n$ the number of generations,
the Ising spin system is magnetized and the magnetization
distribution is non gaussian, with the existence of
local maxima. In the case of XY spins \cite{MelinXY},
the spin system is
magnetized below a cross--over temperature scale $J/n$,
but there are no local maxima in the magnetization distribution,
even though there is a large, non gaussian tail.

The magnetization distribution is investigated using two
techniques. First, 
we calculate analytically the asymptotic
scaling of the moments $\langle M^{q}
\rangle$ for large system sizes, where $M$ is the total
magnetization. We show that in the case
of Ising spins and in the case of XY spins, the
moments of the magnetization have
a non trivial scaling. Below a temperature $T'$ that is defined
in the bulk of the paper, there is one scaling dimension
lower than unity but larger than $1/2$.
If the temperature is larger than $T'$ and lower than the
Bethe-Peierls temperature, there are two scaling dimensions,
but one of them (even moments)
is $1/2$ (scaling dimension of
random walks).
The other one (odd moments)
is lower than $1/2$ but larger than zero. The reason
why the scaling dimension of odd moments is non zero
is that the ancestor's spin is frozen in a given direction.
Above the Bethe--Peierls temperature, all the
spins are uncorrelated and the scaling dimensions
are trivial:
$1/2$ for even moments and zero for odd moments.

Next, we study numerically the fractal properties
of the magnetization distribution measure
in the Ising case.
The magnetization distribution
defines a measure on the interval $[-N_n,N_n]$, where $N_n$
is the number of sites:
\begin{equation}
N_n=1+2+...+2^{n} = 2^{n+1}-1
.
\end{equation}
There exists a temperature regime
where this measure is fractal in the sense
of \cite{multi}. The fractal dimension
does not decay monotonously as the temperature decreases.

We also analyze the overlap distribution and reach similar
conclusions.
\section{Bethe Peierls transition}
\label{BPT}
\subsection{Ising case}
It is know since the thirties \cite{BP} that the Ising
model on the Cayley tree has a mean-field like transition.
For a review on spin systems in the Bethe-Peierls
approach, see \cite{Thorpe} and \cite{Baxter}.
The idea is to decimate spins at the boundary of a
$n$--half--space--tree and to obtain an equivalent
$(n-1)$--half--space--tree with a magnetic field at the
boundary.
We note $Z_n(\beta,h,h_n)$ the partition function of
a $n$--half--space--tree with a magnetic field
$h_n$ at the boundary. We first consider the situation
where two spins $\sigma_1$ and $\sigma_2$
are connected to a common ancestor $\Sigma$. The partition
function is
\begin{equation}
z(\Sigma) = \left(\sum_{\sigma} e^{\beta J \Sigma \sigma}
e^{\beta h_n \sigma} 
\right)^{2}
.
\end{equation}
The summation over $\sigma_1$ and $\sigma_2$ have been
factored out since there are no loops.
We next write $z(\Sigma)$ under the form
$z(\Sigma) = {\cal N} \exp{\left(\beta T.h_n \right)}$.
Since there are two equations (one for $\Sigma=1$ and
one for $\Sigma=-1$), the parameters ${\cal N}$ and
$T.h_n$ exist and are uniquely determined by
\begin{eqnarray}
{\cal N}^{2} &=& z(+)z(-) =
\left( 4 \left( \cosh^{2}{(\beta J)} + \sinh^{2}
{(\beta h_n)} \right)\right)^{2}\\
T.h_n &=& \frac{1}{2 \beta} \ln{\left(\frac{z(+)}
{z(-)} \right)} = \frac{1}{\beta}
\ln{\left( \frac{\cosh{(\beta(J+h_n))}}
{\cosh{(\beta(J-h_n))}} \right)}
.
\end{eqnarray}
The zero field behavior of $T.h$ is
\begin{equation}
T.h = 2 h \tanh{(\beta h)} \equiv p^{(Ising)}(\beta) h
.
\end{equation}
There exists a finite temperature $\beta_{bp}$ such
as $p(\beta_{bp})=1$. If $\beta>\beta_{bp}$,
the magnetic field is exponentially amplified, leading
to a mean field type phase transition in the
Bethe-Peierls limit \cite{Baxter}. If $\beta<\beta_{bp}$, the
magnetic field is irrelevant.
As far as the $n$--half--space--tree is concerned,
we have
\begin{equation}
Z_n(\beta,h,h_n) = \left(4 \left(\cosh^{2}{(\beta J)}
+ \sinh^{2}{(\beta h_n)} \right)\right)^{
\frac{2^{n-1}}{2}}
Z_{n-1}(\beta,h,T.h_n)
.
\end{equation}

\subsection{XY spins}
We now turn to the case of XY spins.
We assume that a small magnetic field $h$ is switched
on at the boundary. The partition function of two spins
connected to a common ancestor $\Sigma$ is
\begin{equation}
\tilde{Z}(\theta) = \left(\int d\theta_1
e^{\beta h \cos{\theta_1}} e^{\beta J \cos(\theta
- \theta_1)} \right)^{2}
\label{ZXY}
.
\end{equation}
Since the magnetic field is small, we can expand
(\ref{ZXY}) up to the second order in $h$ and reexponentiate
to obtain
\begin{equation}
\tilde{Z}(\theta) = \left(1 + \frac{\beta^{2} h^{2}}{2}
\right) \left( \int_0^{2 \pi} du e^{\beta J \cos{u}} \right)^{2}
\exp{\left(\beta h (\cos{\theta})
p^{(XY)}(\beta)\right)}
,
\end{equation}
with
\begin{equation}
p^{(XY)}(\beta) = 2 \int_0^{2 \pi} du \cos{u} w(u)
,
\end{equation}
where $w(u)$ is the probability that the angular difference
between two nearest neighbor spins is $u$:
\begin{equation}
\label{w(u)}
w(u) = \frac{e^{\beta J \cos{u}}}
{ \int_0^{2 \pi} du \exp{(\beta \cos{u})}}
.
\end{equation}
Notice that by replacing $\int d \theta_1$ by
$\int d \theta_1 \left(\delta(\theta_1)
+ \delta(\theta_1-\pi) \right)$, we recover the
Ising case: $w(0)$ is the probability $x=w(0)=e^{-\beta J}/
(e^{\beta J}+ e^{-\beta J})$ that two nearest-neighbor
spins are antiparallel; $w(\pi)=1-x=e^{-\beta J}/
(e^{\beta J}+e^{-\beta J})$ is the probability that
two nearest neighbor spins are parallel; $p^{(XY)}
(\beta)$ becomes $p^{(Ising)}(\beta)$.

\section{Scaling of the moments of the magnetization}
This section is organized as follows: we first derive the
magnetization distribution recursion for Ising spins.
The XY spin case is simply related to the
Ising case by the aforementioned
transformation $\int d\theta
\rightarrow \int d \theta \left( \delta(\theta) +
\delta(\theta-\pi) \right)$. The next step
is to calculate by recursion the scaling behavior
of all the moments.

\subsection{Magnetization distribution recursion}
The aim of this section is to derive the magnetization
distribution recursion for XY spins.
We call $P_n^{\theta}({\bf M})$ the conditional
probability density of the
magnetization of XY spins on a $n$-half-space-tree,
the ancestor pointing in the direction $\theta$.
We put two $n$-half--space--trees together as pictured
on figure \ref{Fig1} to obtain a $(n+1)$--half--space--tree.
Since the link variables are statistically independent, we
have \cite{MelinXY}
\begin{equation}
\label{recM}
P_{n+1}^{\theta}({\bf M}) = \int d {\bf M}_1 d {\bf M}_2
\delta \left(
{\bf M}-{\bf M}_1-{\bf M}_2 - \hat{\theta} \right)
\int d \theta_1 d \theta_2 w(\theta-\theta_1)
w(\theta-\theta_2) P^{\theta_1}_n({\bf M}_1)
P^{\theta_2}_n({\bf M}_2)
,
\end{equation}
where $w(u)$ is defined by (\ref{w(u)}).

\subsection{Lemma}
\label{Lemma}
The aim of this section is to derive recursion relations
for the average of the moments.
We prove the following proposition by induction:
\begin{eqnarray}
{\cal H}^{(1)}_{n-1} &\iff&
\forall q \ge 0,
\langle {\bf M}^{q} \rangle_{n-1}^{\theta} =
\left\{ \begin{array}{cl} A_{n-1}^{(q)}
& \mbox{ if $q$ is even}\\ A_{n-1}^{(q)}
\hat{\theta} & \mbox{if $q$ is odd}
\end{array} \right\},\\
&& \mbox{where $A_k^{(q)}$ is independent on $\theta$}
\nonumber
.
\end{eqnarray}
The average $\langle ... \rangle_n^{\theta}$
means an average over a
$n$--half--space--tree with the spin
ancestor frozen in the direction $\theta$.
A a by-product, we derive a recursion relation for
the $A_n^{(k)}$'s.
Clearly, ${\cal H}^{(1)}_0$ is true,
with $A_0^{(q)}=1$. We now show that ${\cal H}_{n-1}
\Longrightarrow {\cal H}_n$. We first write $\langle
{\bf M}^{q} \rangle^{\theta}_n$ under the form
\begin{equation}
\langle {\bf M}^{q} \rangle_n^{\theta} =
\sum_{l=0}^{q} {q \choose l} 
\sum_{m=0}^{l} {l \choose m} I_{n,q,l,m}(\theta)
,
\end{equation}
where
\begin{equation}
I_{n,q,l,m}(\theta)=\int d \theta_1 d \theta_2
w(\theta-\theta_1) w(\theta-\theta_2)
\left( \langle {\bf M}^{m} \rangle_{n-1}^{\theta_1} .
\langle {\bf M}^{l-m} \rangle_{n-1}^{\theta_2} \right)
.\hat{\theta}^{q-l}
.
\end{equation}
Next, assuming ${\cal H}_{n-1}^{(1)}$,
we consider the eight cases where $q$, $l$ and $m$ are
even or odd, and show that $I_{n,q,l,m}(\theta)$ is either
independent on $\theta$ if $q$ is even or $I_{n,q,l,m}(\theta)=
|I_{n,q,l,m}| \hat{\theta}$, where $|I_{n,q,l,m}|$ is independent
on $\theta$. It would be tedious to enumerate here the
eight cases, and we just pick--up only one case as an exemple:
$q$ odd, $l$ odd and $m$ even. In this case, we have
\begin{equation}
I_{n,q,l,m}(\theta) = \int d \theta_1 d \theta_2 
w(\theta-\theta_1) w(\theta-\theta_1) A_{n-1}^{(m)}
A_{n-1}^{(l-m)} \hat{\theta_2}
.
\end{equation}
Using the fact that $w(-u)=w(u)$,
it is easy to show that
\begin{equation}
\int d \theta_2 w(\theta-\theta_2) \hat{\theta_2}
= \hat{\theta} \int d u w(u) \cos{u}=
\frac{p}{2} \hat{\theta}
,
\end{equation}
which is of the announced form. We have written
$p$ instead of $p^{(XY)}(\beta)$ because
the form of this result is independent on whether
we have XY or Ising spins: one just has to choose
$p^{(Ising)}$ or $p^{(XY)}$ according to the nature
of the spin variables.
We can also show that $I_{n,q,l,m}$ is independent on $q$.
More precisely, we note
\begin{equation}
\gamma_{l,m} = \left\{
\begin{array}{cl}
p/2 & \mbox{if $l$ is odd}\\
1 & \mbox{if $l$ is even and $m$ is even}\\
\left(p/2 \right)^{2} &
\mbox{if $l$ is even and $m$ is odd.}
\end{array}
\right.
\end{equation}
With this notation, it is easy to show that
\begin{equation}
|I_{n,q,l,m}| = A_{n-1}^{(m)} A_{n-1}^{(l-m)} \gamma_{l,m}
,
\end{equation}
which leads us to the following recursion for the $A_n^{(q)}$'s:
\begin{equation}
A_n^{(q)} = \sum_{l=0}^{q} {q \choose l}
\sum_{m=0}^{l} {l \choose m} \gamma_{l,m}
A_{n-1}^{(m)} A_{n-1}^{(l-m)}
(1-\delta_{l,q} \delta_{m,q})
(1-\delta_{l,q} \delta_{m,0})
+ (\gamma_{q,q}+\gamma_{q,0}) A_{n-1}^{(q)}
,
\label{rec}
\end{equation}
where, for further purposes,
we have separated the term $A_{n-1}^{(q)}$
from the terms $A_{n-1}^{(k)}$, with $k<q$.

\subsection{Asymptotic behavior of the moments}
\label{sec3.3}
Our aim is to calculate the asymptotic behavior of the
moments of the magnetization. We have to distinguish between
two cases: $\beta>\beta'$ and $\beta<\beta'$, where
$\beta'$ is defined by $p^{2}=2$. Notice that $\beta'$
corresponds to the inverse temperature above which the
susceptibility per spin diverges \cite{Y}. This is
in fact not surprising since the susceptibility can be calculated
from the moments of the magnetization using the
fluctuation--dissipation theorem.

\subsubsection{Low temperature regime: $\beta>\beta'$}
\label{lowT}
In the regime $\beta>\beta'$, we prove by induction
the following proposition:
\begin{eqnarray}
\nonumber
{\cal H}_{q-1}^{(2)}
&\iff& \mbox{ $\forall k, k\in \langle 0,...,q-1 \rangle$,
the asymptotic behavior of $A_n^{(k)}$}\\
&& \mbox{in the large $n$ limit is
$A_n^{(k)} \sim \alpha_k p^{n k}$, }
\mbox{where $\alpha_k$ is independent on $n$.}
\label{thisprop}
\end{eqnarray}
${\cal H}^{(2)}_{0}$ is obviously true and $\alpha_0=1$.
We show that ${\cal H}^{(2)}_{q-1}
\Longrightarrow {\cal H}^{(2)}_q$, and as a byproduct, we
find the recursion
relations for the $\alpha_q$'s.

We first examine the case where $q$ is odd.
Then, $\gamma_{q,q}+\gamma_{q,0} = p$.
Using (\ref{rec}) we can express
$A_n^{(q)}$ in terms of the $A_i^{(k)}$ with $k<q$
We find
\begin{equation}
\label{sol}
A_n^{q} = p^{n} +
\sum_{l=0}^{q} {q \choose l} \sum_{m=0}^{l}
{l \choose m} \gamma_{l,m} X_{n,l,m}
(1-\delta_{l,q} \delta_{m,q})
(1-\delta_{l,q} \delta_{m,0})
,
\end{equation}
with
\begin{equation}
X_{n,l,m} = \sum_{i=0}^{n-1} p^{n-1-i} A_i^{(m)}
A_i^{(l-m)}
.
\end{equation}
We now assume ${\cal H}_{q-1}^{(2)}$
in order to find the asypmtotic behavior of $X_{n,l,m}$,
in the large $n$ limit, which is
\begin{equation}
X_{n,l,m} \sim \alpha_m \alpha_{l-m} \frac{1}{p^{l}-p}
p^{nl}
.
\end{equation}
The case $q=1$ is special due to the fact that one has to
take into account the $p^{n}$ term in (\ref{sol}). This case
has to be treated separately, and one finds \cite{Melin} \cite{MelinXY}:
$A_n^{(1)} \sim \alpha_1 p^{n}$, with $\alpha_1=p/(p-1)$.
If $q>1$, the $p^{n}$ term in (\ref{sol}) is sub--dominant,
and we get for $q>1$ $A_n^{(q)} \sim \alpha_q p^{qn}$, with
\begin{equation}
\alpha_q = \frac{1}{p^{q}-p}
\sum_{m=1}^{q-1} {q \choose m}
\gamma_{q,m} \alpha_m \alpha_{q-m}
.
\end{equation}
If $q$ is even, $\gamma_{q,q}+\gamma_{q,0}=2$.
We carry out the same reasoning and
we find the following recursion relation for $\alpha_q$
if $q \ge 2$:
\begin{equation}
\alpha_q = \frac{1}{p^{q}-2}
\sum_{m=1}^{q-1} {q \choose m} \gamma_{q,m}
\alpha_m \alpha_{q-m}
.
\end{equation}
We have used the assumption $\beta>\beta'$ (ie $p^{2}>2$)
to determine which term is dominant in the large $n$ limit.

Notice that $\alpha_q$ is effectively independent of $n$, as
was stated in proposition (\ref{thisprop}). In the case $q=2$,
we find that
\begin{equation}
\alpha_2 = \frac{1}{2} \frac{p^{4}}{(p-1)^{2}} \frac{1}{p^{2}-2}
,
\end{equation}
which is exactly what was found in \cite{MelinXY}.
\subsubsection{Intermediate temperature regime: $\beta_{bp}<\beta
<\beta'$}
In the regime $\beta_{bp}<\beta<\beta'$, we call ${\cal H}^{(3)}$
the following proposition:
\begin{eqnarray}
\nonumber
{\cal H}_{q-1}^{(3)} &\iff& \forall k, k \in
\langle 0,...,q-1 \rangle,
\mbox{the asymptotic behavior of $A_n^{(k)}$ in the large $n$ limit
is}\\
&& A_n^{(k)} \sim \left\{
\begin{array}{cl}
\beta_k 2^{n k /2} & \mbox{if $k$ is even}\\
\beta_k \left(2^{(k-1)/2} p \right)^{n} & \mbox{if $k$ is odd}
\end{array}
\right\},
\mbox{where $\beta_k$ is independent on $n$}
,
\end{eqnarray}
and we show by induction that ${\cal H}^{(3)}$ is true.
We have shown in \cite{MelinXY} that ${\cal H}_0^{(3)}$ and
${\cal H}_1^{(3)}$ are true. The technique to show that
${\cal H}_{q-1}^{(3)} \Longrightarrow {\cal H}_q^{(3)}$
is the same
as in section \ref{lowT}: we use (\ref{rec}) and we
analyze separately the case $q$ even and $q$ odd.
After straightforward calculations, we conclude that
${\cal H}_{q-1}^{(3)} \Longrightarrow {\cal H}_q^{(3)}$,
and that
\begin{equation}
\beta_q =
\frac{1}{2^{q/2}-2}
\sum_{\shortstack{ \scriptsize $m=2$\\
\scriptsize $m \equiv 0 (2)$}}^{q-2} {q \choose m}
\gamma_{q,m} \beta_m \beta_{q-m}
\end{equation}
if $q$ is even, and
\begin{equation}
\beta_q = 
\frac{1}{p\left( q^{(q-1)/2}-1 \right)}
\sum_{m=1}^{q-1}  {q \choose m} \gamma_{q,m} \beta_m
\beta_{q-m}
\end{equation}
if $q$ is odd.
\subsection{Discussion}
We have thus shown that if $\beta>\beta'$, the dominant behavior
of the moments of the magnetization in the large $n$ limit is
\begin{equation}
\langle M^{q} \rangle^{\theta}
\propto N^{q \frac{\ln{p}}{\ln{2}}}
,
\end{equation}
where $N$ is the number of sites.
Using the analogy to multifractals
\cite{multi}, this means that
there is one singularity of strength $\alpha=\ln{p}/\ln{2}$,
and of scaling dimension $f=\ln{p}/\ln{2}>1/2$.
Now, if $\beta_{bp} < \beta<\beta'$, we have
\begin{equation}
\langle M^{q} \rangle^{\theta} \propto N^{q/2}
\end{equation}
if $q$ is even and
\begin{equation}
\langle M^{q} \rangle^{\theta}
\propto N^{\frac{q-1}{2} + \frac{\ln{p}}
{\ln{2}}}
\end{equation}
if $q$ is odd. There are two singularities of scaling dimension
$f_1(q) = 1 / 2$ if $q$ is even and $f_2(q) = \ln{p}/\ln{2}$ if
$q$ is odd. If the temperature goes to the Bethe--Peierls
transition temperature, $f_1(q)=1/2$ and $f_1(q)=0$.
At the
Bethe--Peierls transition, the correlation length is shorter
than the lattice spacing \cite{Melin}, and all the
spins are thus uncorrelated. The magnetization is thus
the sum of $N$ random independent spin variables.
Above the Bethe-Peierls temperature, the
scaling dimension
of even moments is $1/2$, as the one of random walks \cite{ID},
and the scaling dimension of odd moments is zero.
Notice that these result are true for XY spins as well as
Ising spins. One just has to pick--up the right $p$ function.
This conclusion is to be contrasted
with the behavior of the magnetization distribution. We have shown
numerically that the magnetization distribution
has local maxima in the case of Ising spins \cite{Melin}, and
no local maxima in the case of XY spins \cite{MelinXY}. Even though
the scaling of the moments of the magnetization of XY spins
is non trivial,
the magnetization distribution is a smooth function, with a long tail,
but obviously no local maxima, unlike the case of Ising spins
where, as explained in section \ref{Mag dis}
the magnetization distribution
has a proliferation of local maxima,
at low temperatures, and we find non trivial fractal
dimensions for the probability distribution measure.

\section{Magnetization distribution in the Ising case}
\label{Mag dis}
The aim of this section is to analyze the low temperature
magnetization distribution $P(M)$ in the Ising case, where
local maxima of the magnetization distribution are present
\cite{Melin}. The main result of this section is the
computation of the fractal dimension of the magnetization
distribution at low temperatures.
By fractal dimension, we mean the fractal dimension
of the measure $P(M)$, as defined in
\cite{multi}.
We insist on the fact
that these results are strongly dependent on the
discrete nature of the spin variables.

It was argued in \cite{Melin} that below a temperature
of the order $J/\ln{n}$, the magnetization is non gaussian.
This temperature scale corresponds to the temperature
cross--over in
\begin{equation}
\frac{\langle M \rangle_n^{+}}{N_n}
= \frac{\alpha_1}{2} \left(\frac{p}{2} \right)^{n}
 \simeq \frac{\alpha_1}{2}\left(1-n e^{-2 \beta J} \right)
,
\end{equation}
which leads to the aforementioned cross--over temperature.
Our goal is to investigate the scaling properties of the low
temperature magnetization distribution.
The magnetization distribution is plotted on figure \ref{Fig2}
for three temperatures. We see that as the temperature decreases,
the magnetization distribution looks more and more fractal.
In order to put this observation on a quantitative basis,
we divide the segment $[-N_n,N_n]$ into
$l$ pieces of size $2^{-l+1} N_n$,and we
call $P_i$ the probability measure of the $i$-th segment.
We calculate \cite{multi}
\begin{equation}
\chi_l(q) = \sum_i P_i^{q}
,
\end{equation}
and look for the behavior of $\chi_l(q)$ for large $l$.
If the temperature is not too low, we can identify
the existence of a scaling regime, namely $\ln{\chi_l(q)}$
is linear as a function of $-l+1$ for large $l$ (see figure
\ref{Fig3}).
In practise, one cannot reach the $l=+\infty$ regime
because the magnetization is a discrete variable.
The next step consists in calculation the slope of $\ln
{\chi_l(q)}$ as a function of $\ln{2^{-l+1}}$ in the
scaling regime:
\begin{equation}
(q-1)D_q = \mbox{slope of } \ln {\chi_l(q)} \mbox{versus }
\ln{2^{-l+1}}=q \alpha - f
,
\end{equation}
where $\alpha$ is the strength of the singularity, $f$
the fractal dimension of the probability measure
and $D_q$ the generalized dimension.
In the multifractal case,
both $\alpha$ and $f$ depend on $q$. In our case, we
find that $(q-1)D_q$ is linear as a function of $q$,
so that there is only one fractal dimension in our problem.
The fractal dimension is plotted on figure \ref{Fig4} as
a function of the inverse temperature for different sizes.
The fractal dimension of the magnetization distribution
does not decay monotonously to zero as the temperature
decreases, and we observe finite size effects.
We have no analytic understanding of the existence
of temperature domains where the fractal dimension
of the magnetization increases as the temperature decreases.
Finite size effects are understood as follows.
Below a temperature $T^{*}$ of the order of $J/\ln{n}$,
the kinks (ie antiparallel nearest neighbor spins)
give rise to well--defined domains of magnetization \cite{Melin}.
Consider one kink on a $n$--half--space--tree. The probability
to find a kink at generation $k$ (counted from the ancestor) is
\begin{equation}
P_n(k)=\frac{2^{k}}{2(2^{n}-1)}
.
\end{equation}
A kink at generation $k$ would involve $2^{n-k+1}-1$ spin flips.
Now consider one kink on a $(n+1)$--half--space--tree. The probability
to find a kink at generation $k+1$ is
\begin{equation}
P_{n+1}(k+1)=\frac{2^{k+1}}{2(2^{n+1}-1)}
,
\end{equation}
which would also involve $2^{n-k+1}-1$ spin flips.
Since, in the limit of large $n$, $P_n(k) \simeq P_{n+1}(k+1)$,
the magnetization distribution is essentially
the same on a $n$--half--space--tree and a $(n+1)$--half--space--tree
provided the temperature is rescaled such as the number of
antiparallel bonds is the same. In other words, we should not plot
the fractal magnetization of the magnetization distribution versus
the inverse temperature, but rather versus
the number of antiparallel bonds $N_b=x N_n$. When the
fractal dimension of the magnetization probability
measure is plotted versus
the number of antiparallel bonds, all the curves collapse on a
single master curve (see figure \ref{Fig5}).

Then, one question arises: numerically, we can only reach a small
number of generations. Is it still a master curve in the
thermodynamic limit? The answer is yes. We showed in \cite{Melin}
that the magnetization distribution is non gaussian below a
temperature scale $T^{*} \sim J / \ln{n}$ which tends slowly
to zero if the number of sites increases. Since we are interested
in the physics with a small number of broken bonds (see figure
\ref{Fig5}), this cross-over does not come into account.
More precisely, we can calculate the number of broken
bonds at the cross--over temperature $T^{*}$:
\begin{equation}
N_b(T^{*}) = N_n \frac{e^{-\beta^{*}J}}{e^{-\beta^{*}J}
+ e^{\beta^{*} J}} \sim \frac{2^{n+1}}{n^{2}+1}
.
\end{equation}
This quantity diverges in the $n \rightarrow + \infty$
limit, so that the master curve (fractal dimension of
the measure $P(M)$ versus the number of broken bonds)
still exists in the thermodynamic limit.
In other words, we are interested here in the physics
at temperatures of the order $J/n$, which decays much faster
than the cross--over scale $T^{*} \sim J / \ln{n}$.

\section{Scaling of the moments of the overlap distribution}
\subsection{Recursion relation for the overlap distribution}
We now wish to analyze the overlap distribution in the same way
as we analyzed the magnetization distribution. From now on,
we work only with Ising spins. The notion of overlap distribution
$P(q)$ was introduced by Parisi in the context of spin glass theory
\cite{Parisi}. The idea is to consider two uncoupled copies of the
same spin system at the same temperature and to define the
overlap as
\begin{equation}
q[\{\sigma_i\},\{\tilde{\sigma}_i\}]
= \frac{Q}{N} = \frac{1}{N} \sum_{i=1}^{N} \sigma_i
\tilde{\sigma}_i
,
\end{equation}
where $\{\sigma_i\}$ and $\{\tilde{\sigma}_i\}$ are the spin variables
in the two copies.
The overlap distribution is given by
\begin{equation}
P(q) = \sum_{\{\sigma_i\}} \sum_{\{\tilde{\sigma}_i\}}
P(\{\sigma_i\}) P(\tilde{\sigma}_i)
\delta(q- q[\{\sigma_i\},\{\tilde{\sigma}_i\}])
.
\end{equation}
More specifically, we call $P_n^{++}(Q)$ the probability that
the overlap between two copies is $Q$, given that the two ancestor's
spins are $+$. Using the fact that the link variables are statistically
independent, we have
\begin{eqnarray}
\label{recQ}
P_{n+1}^{++}(Q) &=& \sum_{Q_1} \sum_{Q_2}
\left[ \alpha P_n^{++}(Q_1) P_n^{++}(Q_2)
+ \beta P_n^{++}(Q_1) P_n^{++}(-Q_2) \right.\\
&&
+ \left. \gamma P_n^{++}(-Q_1)  P_n^{++}(-Q_2) \right]
\delta(Q-Q_1-Q_2-1)
\nonumber
,
\end{eqnarray}
where $\alpha$, $\beta$ and $\gamma$ are
some temperature dependent coefficients:
$\alpha=(1-x)^{4}+2 x^{2}(1-x)^{2} + x^{4}$,
$\beta=4x(1-x)^{3} + 4x^{3} (1-x)$ and
$\gamma=4 x^{2}(1-x)^{2}$.
We have used the fact that $P^{+-}(Q)=P^{++}(-Q)=P^{-+}(Q)$
and $P^{++}(Q)=P^{--}(Q)$.
The recursion relation for the overlap distribution (\ref{recQ}) is
similar to the recursion relation for the magnetization distribution
(\ref{recM}), except that the coefficients are not the same.
We thus expect the same physics, namely the appearance of
non trivial scaling dimensions in the scaling
of the moments of the overlap
distribution as a function of the system size, and also the
fact that the overlap distribution is a fractal measure
at low temperatures.
In what follows, we denote by $B_n^{(q)}$ the $q$-th moment of
the overlap distribution:
$B_n^{(q)}=\langle Q^{q} \rangle^{++}$.

\subsection{First and second moments}
We want to calculate the asymptotic behavior of $B_n^{(1)}$
and $B_n^{(2)}$
in the large $n$ limit. The calculations are straightforward,
so that we do not give the details here.
As far as $B_n^{(1)}$ is concerned,
we obtain two different regimes: if $T<T'$,
\begin{equation}
B_n^{(1)} \sim \frac{p^{2}}{p^{2}-2}
\left(\frac{p^{2}}{2} \right)^{2}
,
\end{equation}
and if $T>T'$:
\begin{equation}
B_n^{(1)} \sim \frac{2}{2-p^{2}}
.
\end{equation}
As far as $B_n^{(2)}$ is concerned, we need to define the
temperature $T''$ such that $p^{4}=8$. We find that
if $T<T''$,
\begin{equation}
B_n^{(2)} \sim \frac{1}{2} \left(\frac{p^{2}}{p^{2}-2} \right)^{2}
\frac{p^{4}}{p^{4}-8} \left( \frac{p^{2}}{2} \right)^{2n}
.
\end{equation}
If $T>T''$,
\begin{equation}
B_n^{(2)} \sim \frac{p^{4}-4 p^{2}+8}{(2-p^{2})^{2}} 2^{n}
.
\end{equation}
\subsection{Lemma}
Carrying out the same calculation as in section \ref{Lemma},
we find that
\begin{equation}
B_n^{(q)} = \sum_{l=1}^{q} {q \choose l}
\sum_{m=0}^{l} {l \choose m} \lambda_{l,m}
(1-\delta_{l,q}\delta_{m,q})(1-\delta_{l,q}
\delta_{m,0}) \sum_{i=0}^{n-1} \mu_q^{n-1-i}
B_i^{(m)} B_i^{(l-m)} + \mu_q^{n}
\label{recB}
,
\end{equation}
where $\lambda_{l,m}=\alpha + (-1)^{m} \beta +
(-1)^{l} \gamma$ and $\mu_q=\lambda_{q,q}+\lambda_{q,0}$.
If $q$ is even, $\mu_q=2$ and $\mu_q=p^{2}/2$ if $q$ is odd.
\subsection{Asymptotic behavior of the moments}
Using the relation (\ref{recB}), we can derive the asymptotic
behavior of the moments.
The idea is to prove the following results
by induction. Using a method similar to the one of section
\ref{sec3.3}, we derive the asymptotic behavior
of $B_n^{(q)}$.

\subsubsection{Low temperature regime: $\beta>\beta''$}
In this regime, the asymptotic behavior of the moments of
the overlap distribution is
\begin{equation}
B_n^{(q)} \sim \gamma_q \left( \frac{p^{2}}{2} \right)^{q n/2}
,
\end{equation}
with the following recursion relation for the $\gamma_q$'s:
\begin{equation}
\gamma_q =
\frac{1}{(p^{2}/2)^{q/2}-2}
\sum_{m=1}^{q-1} {q \choose m} \lambda_{q,m}
\gamma_m \gamma_{q-m}
\end{equation}
if $q$ is even. If $q$ is odd, we have
\begin{equation}
\gamma_q = \frac{1}{(p^{2}/2)^{q/2}-p^{2}/2}
\sum_{m=1}^{q-1} {q \choose m} \lambda_{q,m} \gamma_m
\gamma_{q-m}
.
\end{equation}
\subsubsection{Intermediate temperature regime: $\beta'<\beta<\beta''$}
The asymptotic behavior of the moments in this
temperature regime is
\begin{equation}
B_n^{(q)} \sim \left\{
\begin{array}{cl}
\epsilon_q 2^{qn/2} & \mbox{if $q$ is even}\\
\epsilon_q \left( 2^{(q-3)/2} p^{2} \right)^{n}
& \mbox{if $q$ is odd}
\end{array}
\right.
\end{equation}
The recursion relations for the $\epsilon_q$'s are as follows.
If $q$ is even
\begin{equation}
\epsilon_q = \frac{1}{2^{q/2}-2}
\sum_{\shortstack{ \scriptsize $m=2$\\
\scriptsize $m \equiv 0 (2)$}}^{q-2}
{q \choose m} \lambda_{q,m} \epsilon_m
\epsilon_{q-m}
.
\end{equation}
If $q$ is odd,
\begin{equation}
\epsilon_q = \frac{2}{p^{2} (2^{(q-1)/2}-1)}
\sum_{m=1}^{q-1} {q \choose m} \lambda_{q,m}
\epsilon_m \epsilon_{l-m}
.
\end{equation}
\subsection{Discussion}
We have thus shown that if $\beta>\beta''$,
\begin{equation}
\langle Q^{q} \rangle^{++} \sim N^{q
\frac{\ln{(p^{2}/2)}}{2 \ln{2}}}
,
\end{equation}
with $N$ the number of sites, which leads to
a scaling dimension $f=\ln{(p^{2}/2)}/(2 \ln{2})$.

If $\beta'<\beta<\beta''$,
\begin{equation}
\langle Q^{q} \rangle^{++} \sim N^{q/2}
\end{equation}
for even moments,
corresponding to a scaling dimension $1/2$.
The scaling of the odd moments is
\begin{equation}
\langle Q^{q} \rangle^{++} \sim N^{\frac{q-1}{2}
+ 2 \frac{\ln{p}}{\ln{2}}-1}
,
\end{equation}
leading to a scaling dimension $f=2 \ln{p}/\ln{2}-1$.
If $\beta>\beta''$, the scaling dimensions are trivial
in the sense that the scaling dimension of even momenta is
$1/2$ and the scaling dimension of odd momenta is zero.
Since $\beta'>\beta_{bp}$, there exists a regime of
temperature $\beta'>\beta>\beta_{bp}$ in which the
scaling dimension of the even momenta of both the
magnetization and the overlap are $1/2$, but where
the odd momenta of the magnetization are not trivial
(between $0$ and $1/2$) while the odd momenta of the
overlap distribution are trivial (zero).
\section{Overlap distribution}
We now analyze the overlap distribution $P(Q)$, which is plotted
on figure \ref{Fig6} for different temperatures.
If the temperature decreases, the overlap distribution has local
maxima, and, if the temperature further decreases, the
number of local maxima increases, and at low temperatures, the
overlap distribution looks fractal. Using the technique described
in section \ref{Mag dis}, we can calculate the fractal dimension
of the measure $P(Q)$. This quantity is plotted on figure
\ref{Fig7} as a function of the average number of antiparallel
bonds. Again, we find that the variations of the fractal
dimension of the measure $P(Q)$ as a function of the number of
antiparallel bonds is non monotonous.
\section{Conclusion}
We have analyzed the asymptotic scaling of all the moments
and the fractal dimension of the measures $P(M)$ and $P(Q)$.
There exists a temperature region $\beta'>\beta>\beta_{bp}$
where the magnetization moments are in an intermediate regime
(scaling dimension $1/2$ for even moments and non zero scaling
dimension for odd moments) whereas the moments of the
overlap distribution are in a high temperature regime.
Another conclusion is that in both cases, the fractal
dimension of the measures $P(M)$ and $P(Q)$ do not decrease
monotonously as a function of temperature.
We have no qualitative understanding of this observation.

Two directions for future work are under investigation:
the analysis of fractals and percolation clusters,
where it was shown that the magnetization distribution
has also local maxima at low temperatures \cite{fractal}.
Another direction is the analysis of magnetized spin glasses. It was shown
in \cite{Chayes} that, in the magnetized spin glass phase of
the $\pm J$ model, the magnetization distribution has local
maxima. It would be interesting to compare the results
of the present paper with the behavior of the magnetized spin glass.

I acknowledge Y. Leroyer for his suggestion
to study the fractal scaling of the magnetization
distribution measure. I also acknowledge a discussion with
C. Tang. P. Chandra is acknowledged for her comments
on the manuscript.


\newpage
\renewcommand\textfraction{0}
\renewcommand
\floatpagefraction{0}
\noindent {\bf Figure captions}

\begin{figure}[h]
\caption{}
\label{Fig1}
The way two $4$--half--space--trees are put together to
make a $5$--half--space--tree.
\end{figure}

\begin{figure}[h]
\caption{}
\label{Fig2}
Magnetization distribution of a $n=15$ generations tree,
for $\beta=4.16$, $\beta=3$ and $\beta=1$.
\end{figure}

\begin{figure}[h]
\caption{}
\label{Fig3}
$\chi_l(q)$ as a function of $(-l+1)\ln{2}$ for a $n=15$ half--
space--tree and $\beta=4.9$. We see the existence of a scaling
regime for large $l$.
\end{figure}

\begin{figure}[h]
\caption{}
\label{Fig4}
Fractal dimension of $P(M)$ as a function of the
inverse temperature for $n=8,9,10,11,12,13,14,15$ generations.
\end{figure}

\begin{figure}[h]
\caption{}
\label{Fig5}
Fractal dimension of $P(M)$ as a function of the
average number of antiparallel bonds.
The curves for $n=8,9,10,11,12,13,14,15$ all collapse
on a universal curve.
\end{figure}

\begin{figure}[h]
\caption{}
\label{Fig6}
Overlap distribution of a $n=15$ generations tree,
for $\beta=5.28$, $\beta=3$ and $\beta=1$.
\end{figure}

\begin{figure}[h]
\caption{}
\label{Fig7}
Fractal dimension of $P(Q)$ as a function of the
average number of antiparallel bonds.
The curves for $n=8,9,10,11,12,13,14,15$ all collapse
on a universal curve.
\end{figure}
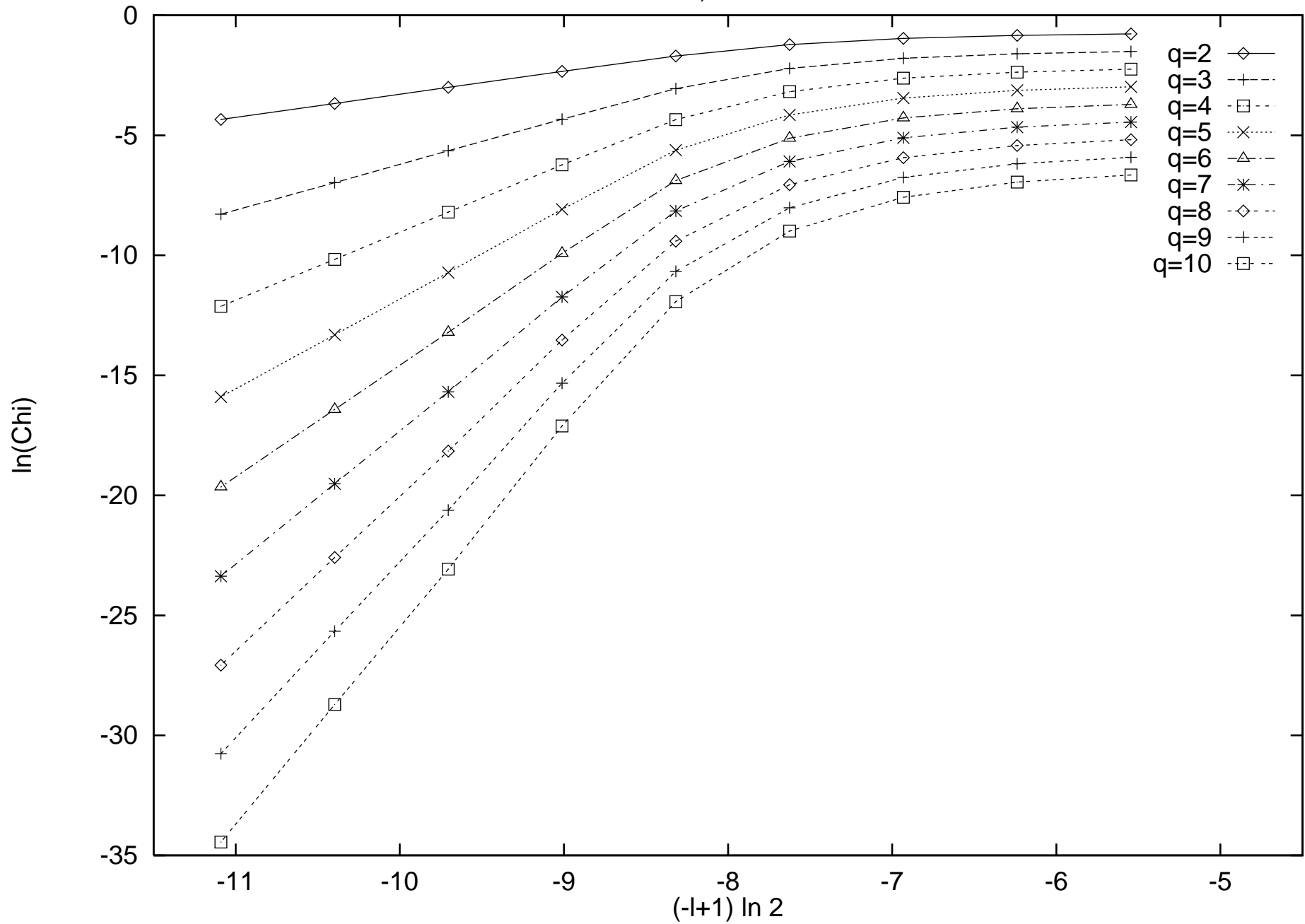
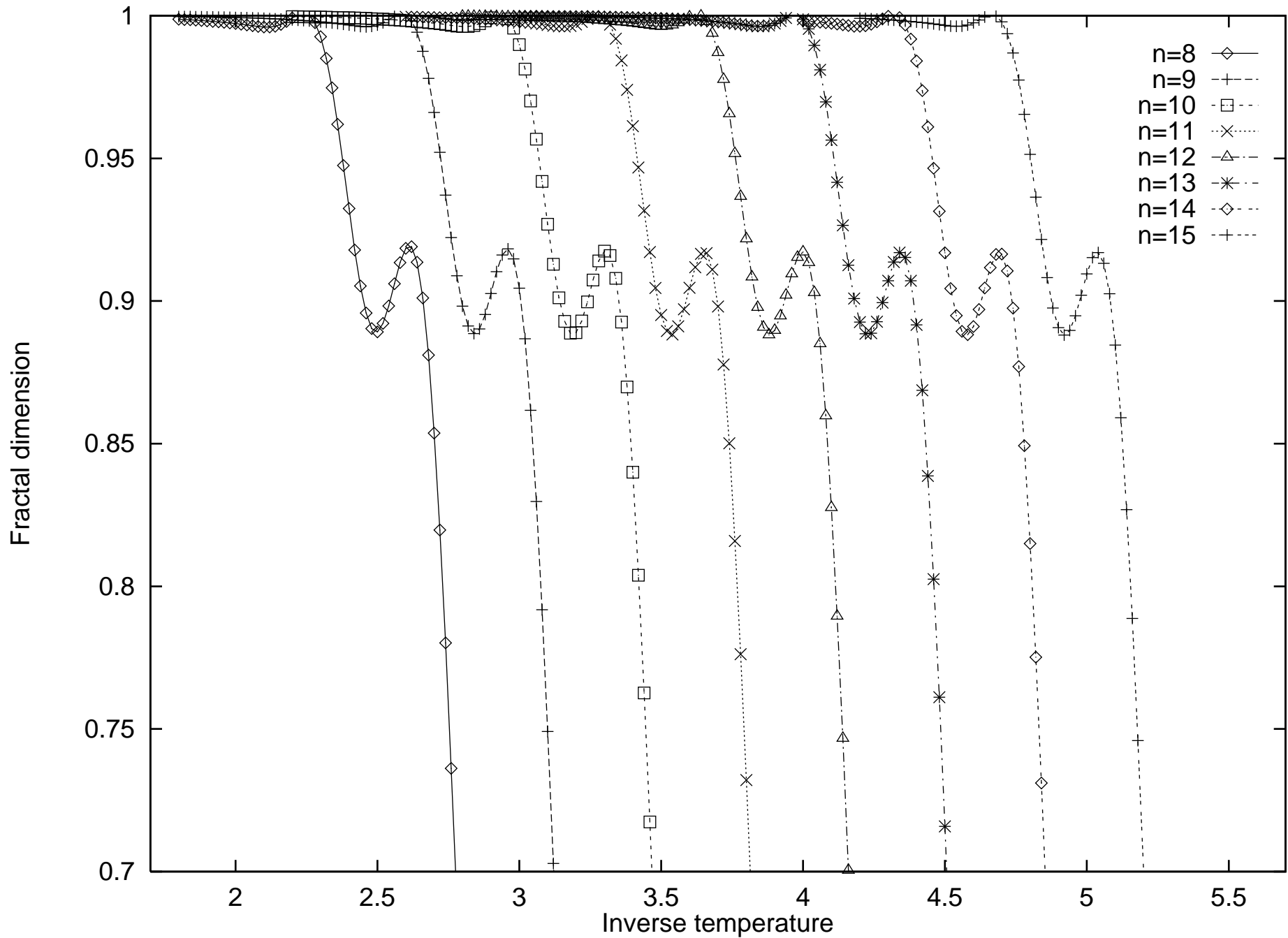
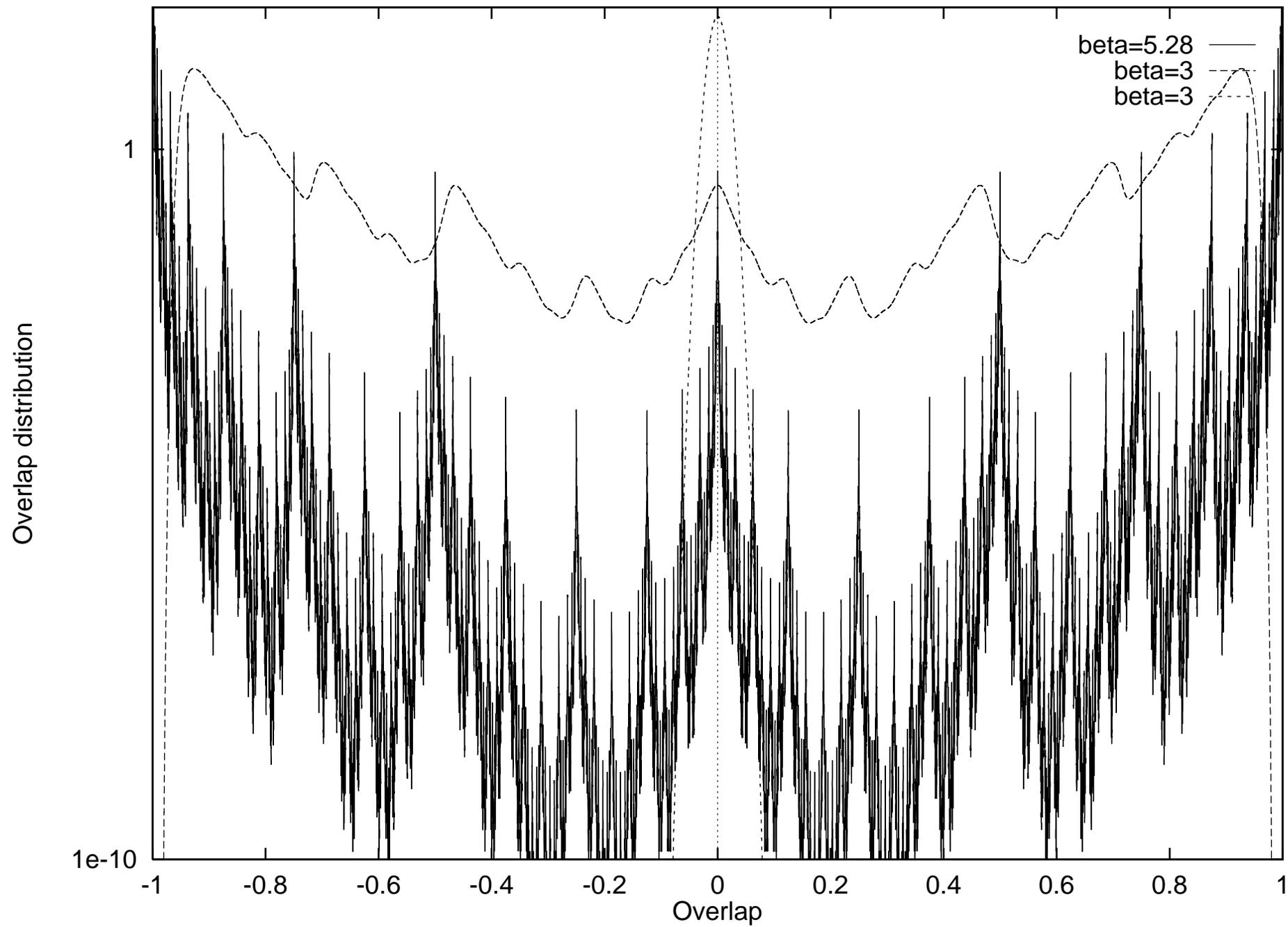
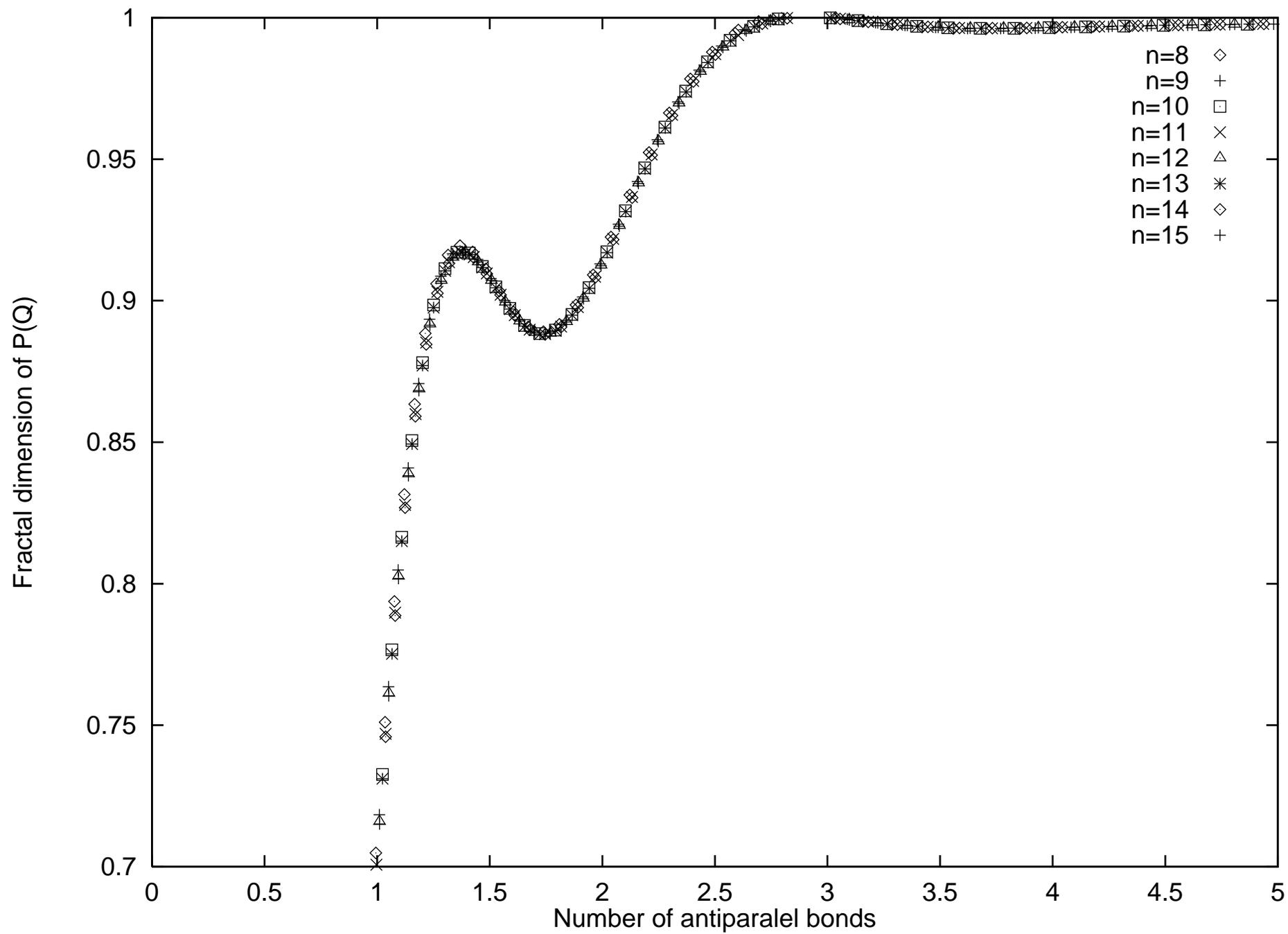
\end{document}